\documentclass[conference]{IEEEtran}
\IEEEoverridecommandlockouts
\usepackage{cite}
\usepackage{amsmath,amssymb,amsfonts}
\usepackage{algorithmic}
\usepackage{graphicx}
\usepackage{textcomp}
\usepackage{xcolor}
\def\BibTeX{{\rm B\kern-.05em{\sc i\kern-.025em b}\kern-.08em
    T\kern-.1667em\lower.7ex\hbox{E}\kern-.125emX}}
\begin{document}

\title{Intelligence of Things: A Spatial Context-Aware Control System for Smart Devices}

\author{
\IEEEauthorblockN{
Sukanth Kalivarathan\IEEEauthorrefmark{1}, 
Muhmmad Abrar Raja Mohamed\IEEEauthorrefmark{2}, 
Aswathy Ravikumar\IEEEauthorrefmark{3}, 
S Harini\IEEEauthorrefmark{4}
}
\IEEEauthorblockA{
\IEEEauthorrefmark{1}\IEEEauthorrefmark{2}Student, School of Computer Science and Engineering,\\
Vellore Institute of Technology, Chennai, India \\
\IEEEauthorrefmark{1}sukanth.k2021@vitstudent.ac.in, 
\IEEEauthorrefmark{2}muhmmadabrar.r2021@vitstudent.ac.in
}
\IEEEauthorblockA{
\IEEEauthorrefmark{3}AI/ML Consultant, Sustainable Living Labs, India \\
aswathyravi2290@gmail.com
}
\IEEEauthorblockA{
\IEEEauthorrefmark{4}Head of Department \& Professor, School of Computer Science and Engineering,\\
Vellore Institute of Technology, Chennai, India \\
harini.s@vit.ac.in
}
}

\maketitle

\begin{abstract}
This paper introduces Intelligence of Things (INOT), a novel spatial context-aware control system that enhances smart home automation through intuitive spatial reasoning. Current smart home systems largely rely on device-specific identifiers, limiting user interaction to explicit naming conventions rather than natural spatial references. INOT addresses this limitation through a modular architecture that integrates Vision Language Models with IoT control systems to enable natural language commands with spatial context (e.g., "turn on the light near the window"). The system comprises key components including an Onboarding Inference Engine, Zero-Shot Device Detection, Spatial Topology Inference, and Intent-Based Command Synthesis. A comprehensive user study with 15 participants demonstrated INOT's significant advantages over conventional systems like Google Home Assistant, with users reporting reduced cognitive workload (NASA-TLX scores decreased by an average of 13.17 points), higher ease-of-use ratings, and stronger preference (14 out of 15 participants). By eliminating the need to memorize device identifiers and enabling context-aware spatial commands, INOT represents a significant advancement in creating more intuitive and accessible smart home control systems.
\end{abstract}

\begin{IEEEkeywords}
Human-Computer Interaction, Smart Homes, Internet of Things (IoT), Spatial Computing, Vision-Language Models, Natural Language Interfaces, Context-Aware Systems, Multimodal Interaction
\end{IEEEkeywords}

\textbf{}

\begin{figure*}[ht]
    \centering
    \includegraphics[width=\linewidth]{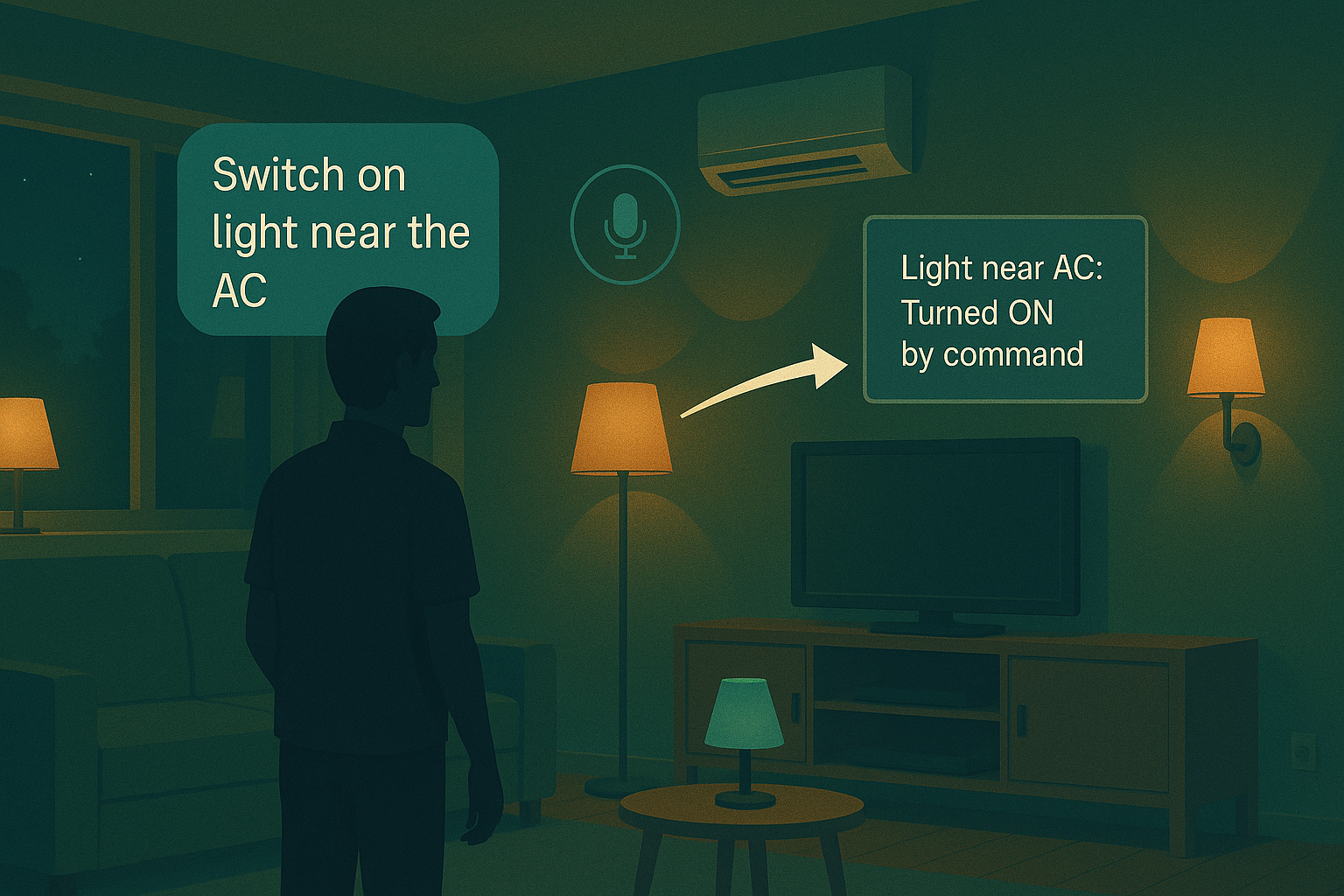}
    \caption{Context-Aware Smart Environment Responding to a Spatially Referenced Voice Command
}
    \label{fig:enter-label}
\end{figure*}

\section{Introduction}

The Internet of Things (IoT) has become a transformative domain in recent years, with a growing emphasis on user-centric design and seamless interaction. Among its many applications, smart homes have seen significant advancements—offering increased functionality, reduced latency, and enhanced automation. However, despite these technical strides, a critical gap remains in how these systems interpret and respond to the context of user commands. Current systems are heavily reliant on voice commands and predefined rule-based automation, often lacking deeper contextual understanding—particularly \textit{spatial context}—within dynamic environments.

Most existing IoT control systems operate through device-specific identifiers such as names or UUIDs. While this approach may suffice for permanent occupants of a household, it introduces considerable friction in public or semi-public spaces such as restaurants, hotels, or assisted living facilities, or when visiting someone else's home where users frequently change. In such scenarios, requiring users to remember or configure specific device names becomes impractical and unintuitive, leading to command ambiguity and reduced accessibility.

There is limited research integrating computer vision with IoT control to improve contextual awareness. Mainstream smart assistants like Google Home and Alexa primarily rely on grouping devices through predefined names or rooms. For example, a user might say, ``Turn on the lights in the living room,'' which works because the lights have been manually grouped. However, if a user wants to control a specific device—say, \textit{``the light near the AC''}—they must know the exact device name or identifier. Ideally, the system should be intelligent enough to understand spatial relationships and resolve such commands naturally, without explicit device names.

Additionally, to emulate human-like spatial reasoning in determining whether to switch a device on or off based on user intent, it is essential to consider how humans typically approach such decisions. Humans assess the purpose of use, identify the appropriate device, and then select the relevant device from among several available options. This nuanced reasoning process is largely absent in current commercial IoT systems and agentic IoT frameworks. Our proposed system seeks to replicate this form of reasoning by leveraging recent advancements in large language models, enabling more context-aware and intent-driven device control. This highlights the need for more \textit{intuitive, spatially-aware interactions} within smart environments.

To address these limitations, we propose \textbf{Intelligence of Things (INOT)} — a spatial context-aware control system that fuses computer vision, natural language processing, and IoT control. INOT’s core innovation lies in its ability to perceive and reason about the physical layout of a smart environment. During a one-time onboarding process, INOT uses zero-shot object detection powered by \textbf{Owl-ViT 2}, a vision-language model, to identify and annotate smart devices within a scene. These annotations are confirmed by the user and passed to \textbf{GPT-4o}, which generates rich spatial context representations. This spatial understanding then acts as the contextual window for downstream agentic execution, driven by \textbf{Gemini Flash}, a lightweight large language model (LLM). Users can then issue commands in natural language—such as \textit{``Turn on the light near the AC''}—and INOT leverages spatial cues to resolve and execute the correct device action.

One of the key strengths of INOT lies in its compatibility with existing IoT ecosystems, enabling easy deployment in real-world environments without requiring extensive hardware modification or software reconfiguration. INOT integrates directly with \textbf{Tuya Smart Device APIs}, a widely-used IoT platform that provides unified access to a diverse range of smart home devices—including lights, fans, thermostats, and appliances—from different manufacturers. This allows INOT to operate across heterogeneous device landscapes without being constrained by vendor-specific protocols.

The onboarding process is designed to be intuitive and requires minimal manual intervention. Upon setup, INOT performs zero-shot device detection using Owl-ViT 2, identifying smart devices in a visual scene without requiring pre-registered device templates. These detections are mapped to Tuya device identifiers through a backend linking process that associates visual entities with the device registry maintained on the Tuya cloud. Once linked and confirmed by the user, INOT can perform real-time control and querying of device states through Tuya's APIs, ensuring smooth two-way communication and status updates.

Moreover, INOT's architecture supports extensibility. It can be adapted to other IoT backends (e.g., Home Assistant, Matter, or proprietary systems) by abstracting device control through modular service adapters. This makes INOT not just a smart home assistant, but a \textit{platform-agnostic control framework} that can scale from single-room installations to enterprise-level environments like smart offices, hotels, and assisted living facilities.

We begin in Section~\ref{sec:relatedwork} by reviewing related work in the fields of IoT device control, smart assistants, and the integration of computer vision in ambient computing. This section highlights the current dependence on device-specific identifiers and voice-only interfaces, drawing attention to the lack of spatial and contextual intelligence in mainstream solutions. We also examine existing efforts to combine vision and language models for object recognition but point out their limited application in real-world IoT ecosystems.

Section~\ref{sec:methodology}, the \textit{Methodology}, describes the core technical architecture of INOT and its seamless integration with real-world IoT platforms such as Tuya. We outline the one-time onboarding pipeline using Owl-ViT 2 for zero-shot visual detection of smart devices, followed by user confirmation of device annotations. The annotated scenes are processed by GPT-4o to generate spatial context, which then serves as the foundation for intelligent agentic execution using Gemini Flash. We explain how user commands in natural language—such as \textit{``Turn on the light near the AC''}—are resolved through spatial reasoning rather than requiring rigid device names. We also discuss how the Tuya API backend enables flexible and vendor-agnostic device control, and how INOT’s architecture supports multilingual input and modular scalability to other smart home ecosystems.

In Section~\ref{sec:userstudy}, we present the \textit{User Study} conducted to evaluate the usability and effectiveness of INOT compared to Google Home Assistant. The study involved 15 participants aged 18–80, with varied educational backgrounds and limited prior experience with smart home technologies. Participants were asked to complete a set of open-ended daily smart home tasks using both systems. Results showed a clear preference for INOT, which was favored for its spatial context awareness, ease of use, and support for regional languages. Quantitative results from the \textbf{NASA-TLX} workload assessment demonstrated a significant reduction in cognitive load when using INOT, affirming the system’s intuitive design and real-world applicability.

Section~\ref{sec:applications} outlines the \textit{practical applications and future directions} of INOT, emphasizing its adaptability across various environments. These include smart glasses with \textit{Simultaneous Localization and Mapping (SLAM)} for hands-free, spatial interaction; static camera-based setups in homes; and assisted living facilities, where intuitive, spatially-grounded commands reduce cognitive and physical effort. The system's effectiveness in elderly care and support for users with cognitive or motor impairments is reinforced by user study results, showing reduced workload and high user satisfaction—especially for those with no prior smart home experience.

Section~\ref{sec:discussions} reflects on the developmental journey of INOT, tracing its evolution from a rule-based prototype in Gazebo to a robust vision-language model-powered system. It discusses key engineering decisions such as moving from manual device annotation to automated recognition and the performance gains from switching to VLMs for spatial reasoning. The section also addresses critical \textit{privacy considerations}, especially in the context of always-on cameras, proposing solutions like on-device inference and encrypted pipelines. Finally, it outlines future directions including user-centric adaptation, such as behavioral modeling, emotional state detection, and federated learning to enable personalization without compromising user data. These insights highlight both the system's current strengths and the roadmap toward more ethical, adaptive, and intelligent IoT interaction.

\section{Previous Works}
\label{sec:relatedwork}
\subsection{Spatial Topology Inference Methods}
Spatial reasoning is a fundamental aspect of artificial intelligence-driven smart home automation, enabling intelligent systems to interpret and respond to their physical surroundings. Existing research has explored various methodologies for spatial understanding, primarily focusing on scene analysis, visual question answering (VQA), and industrial spatial intelligence. However, these approaches primarily serve analytical purposes rather than facilitating real-time automation in universal room environments. Several studies [1-7] have sought to enhance spatial reasoning capabilities in artificial intelligence systems. The ROOT VLM-based scene understanding system \cite{wang2024root} employs an iterative object perception algorithm to detect and annotate objects within indoor environments. While effective in generating structured spatial representations, its primary function remains scene understanding rather than direct automation. Similarly, Spatial VLM \cite{chen2024spatialvlm} introduces large-scale 3D spatial reasoning by training vision-language models with Internet-scale datasets. This approach significantly enhances VQA applications and robotics but does not address dynamic, real-time adaptability for smart home automation.

Further advancements in spatial reasoning have been demonstrated through spatial relations modeling in vision-language systems \cite{yang2024improving}, which introduces methods such as object position regression and spatial relation classification. While these methods refine visual commonsense reasoning tasks, they primarily target improved model performance in VQA and vision-language understanding, rather than practical automation applications. Likewise, industrial spatial intelligence research \cite{wang4926945vlm}, particularly in generation of scene graphs for manufacturing environments, has focused on improving machine perception in structured, predefined industrial settings. While these approaches excel in static, controlled conditions, their adaptability to dynamic, real-world smart home environments remains limited.

Another pertinent study on intelligent control systems for IoT\cite{li2022intelligent} introduces context-aware automation through object-attribute graph modeling. Although this method enables limited automation, it relies primarily on sensor-based data rather than computer vision and lacks robust natural language spatial understanding capabilities.

In a 2024 study Iot-llm \cite{an2024iot} highlight the significant role of Large Language Models (LLMs) in enhancing task reasoning in real-world Internet of Things (IoT) applications, particularly in areas like Human Sensing and Indoor Localization. While this work enhances sensor data interpretation, it lacks vision-language orchestration for environmental-aware automation – resolving device ambiguity through spatial relationships rather than manual labeling or pure sensor fusion.
The most recent work, SAGE \cite{rivkin2023sage} uses LLMs in a fixed prompt tree to process smart home requests, relying on static pre-registered photos for device disambiguation with manually updated relationships. In contrast, our system employs computer vision and Vision-Language Models to dynamically understand wide-scale environments, enabling spatially aware, intuitive commands referencing both smart and non-smart devices, while also tracking those over time.

\subsection{Spatially Aware Smart Home Automation}

The most relevant prior works is a patented smart home system (US9614690) \cite{smartHome2017} which integrates multi-modal inputs, including voice, text, medical sensors, and optical sensors, to facilitate smart home interactions while utilizing a rule-based engine However, its reliance on predefined rules and manually labeled device locations limits its ability to dynamically respond to spatial cues in the physical environment. CANVAS \cite{alrumayh2020context} addresses the complexities of access control in shared smart home environments. By leveraging contextual cues such as user identity, location, and activity, it dynamically adjusts permissions for voice-controlled devices. While this approach enhances privacy, it is primarily limited to access control and lacks a broader automation framework. AutoIoT \cite{cheng2024autoiot} introduces an LLM-powered automation platform capable of generating custom rules using multimodal inputs, including device images and user preferences. It features a conflict detection mechanism that verifies automation rules using formal methods. Although highly flexible, AutoIoT focuses mainly on rule synthesis and does not integrate vision-based device recognition or spatial inference. The study \cite{kok2024iot} explores the integration challenges of Large Language Models (LLMs) in constrained IoT environments. They highlight the need for lightweight architectures and propose techniques such as model compression, task partitioning, and sparse activation to optimize LLM performance on edge devices. However, their work is theoretical and lacks an end-to-end implementation for smart home automation.
The absence of spatial awareness in prior works limits their adaptability. Moreover, while prior work requires explicit user-defined labels for device recognition, our approach facilitates automated detection and contextual interaction.

\subsection{Device Onboarding and Management}
Prior research has laid the foundation for efficient device detection and interaction. The AIDE \cite{zhang2019aide} system provides an augmented onboarding experience by leveraging received signal strength (RSS) profiles to map physical devices to digital identities. However, it lacks a deep contextual understanding of the environment and does not integrate user intent. Unlike AIDE, our system supports multi-modal input, enhances flexibility, and improves accuracy by integrating large language models for structured data extraction. The patent CN109491263B \cite{cnSmartHome2023} describes an image-based control mechanism that maps device positions to interactive control interfaces but does not leverage voice based systems or generative AI for controlling.
While prior works such as \cite{rivkin2023sage} and \cite{cheng2024autoiot} have explored IoT device onboarding for enabling spatial awareness, Meyuhas et al. (2024) \cite{meyuhas2024iot} propose a novel hybrid labeling approach combining string-matching for vendor identification and a Roberta-based model for function labeling. While Meyuhas et al. optimize network-based device labeling, our approach uses computer vision techniques to identify, label and map the devices.

\subsection{Multimodal IoT Systems: Advancements in Spatially-Aware Automation}
Researchers leveraging GPT-3 for contextual smart home control. This study explores the use of large language models (LLMs) to interpret high-level user commands \cite{king2023get} (e.g., "get ready for a party") and map them to specific device actions. The system demonstrates the feasibility of intent-based automation by parsing textual user inputs and generating machine-parseable smart home commands. Limitations include that the system lacks real-time visual scene understanding to validate device placement and state. There is no consideration of spatial relationships between IoT devices, leading to possible misinterpretations. Commands rely primarily on linguistic cues rather than a multimodal fusion of voice, vision, and context.

The system \cite{usMultiModal2017} introduces "voice-clicking," a technique that synchronizes voice commands with non-verbal actions (e.g., touching a screen or making gestures). The goal is to improve context-aware automation by fusing audio and touch data streams. Limitations include that the approach lacks spatial intelligence, limiting its adaptability to dynamic environments. There is no integration with vision-language models (VLMs) for real-time object detection. Device interactions remain tied to predefined user interactions rather than an autonomously inferred scene understanding.

Recent work by Xiao et al. \cite{xiao2024efficient} highlights the role of prompt engineering in enhancing LLM performance for generative tasks in IoT systems. Their approach demonstrates that structured prompts—featuring explicit feature references and standardized output formats—significantly improve the relevance and accuracy of LLM-generated insights from real-time IoT data. While Xiao et al. optimize prompt structure for inference quality, our system emphasizes end-to-end automation, where prompts are dynamically generated and adapted based on user input and spatial metadata.

Zong et al. \cite{zong2025integrating} demonstrate the transformative potential of LLMs in IoT ecosystems, showcasing their ability to interpret complex data streams, enable predictive maintenance, and support natural language interaction for intuitive device control. Their case studies emphasize prompt engineering and cross-device communication to enhance automation and user experience.

\subsection{Agentic IoT Automation Systems}
Prior research in the domain, such as \cite{an2024iot}, \cite{kok2024iot}, \cite{rivkin2023sage}, \cite{zhang2019aide}, \cite{xiao2024efficient}, \cite{zong2025integrating}, \cite{cheng2024autoiot} has explored various methodologies for enhancing smart home automation. Prior works primarily rely on predefined control scripts or hierarchical command execution frameworks to process user intent and control IoT devices. For instance, LLMind structures task execution via finite-state machine (FSM) transformations, ensuring a controlled but rigid flow of automation. While AIoT Smart Home enables smart home control through dynamically constructed LLM-driven decision trees, it lacks the ability to process advanced spatial dependencies.
A critical limitation of previous approaches is their inability to dynamically adapt to real-world changes in device positioning or room configurations. The current invention addresses this by employing an adaptive scene understanding mechanism that periodically updates the object placements and contextual dependencies. Unlike prior works, which treat IoT automation as a static process, this system continuously refines its spatial understanding to enhance automation reliability and efficiency.

\subsection{Applications and Use Cases of IoT Systems with Spatial Reasoning}
Dementia care necessitates continuous monitoring and adaptive interventions to ensure the safety and well-being of persons with dementia (PwDs). Traditional smart home systems, while beneficial, often fail to address the specific needs of PwDs, as they assume a uniform approach to automation \cite{alexakis2019control}. IoT systems plays a pivotal role in AAL by enabling real-time data collection and automation to support elderly individuals \cite{amiribesheli2018tailored}.  The integration of AI-driven spatial reasoning can significantly enhance AAL solutions by Improving Context-Aware Decision Making, Enabling Multimodal Control Mechanisms and Closed-Loop Feedback Mechanism. Despite advancements in IoT and smart home technologies, traditional AAL systems face significant challenges that hinder their widespread adoption \cite{dohr2010internet}. These challenges include Lack of Personalization, Insufficient Social Integration and Limited Integration of Heterogeneous Data Sources. To address these challenges, AI-powered IoT systems must incorporate spatial reasoning capabilities to enhance adaptability, personalization, and real-time responsiveness. Such systems offer Adaptive Automation, Enhanced Human-Agent Interaction and Integration of Multimodal Data Sources.

\section{Methodology}
\label{sec:methodology}
\begin{figure*}[ht]
    \centering
    \includegraphics[width=\textwidth]{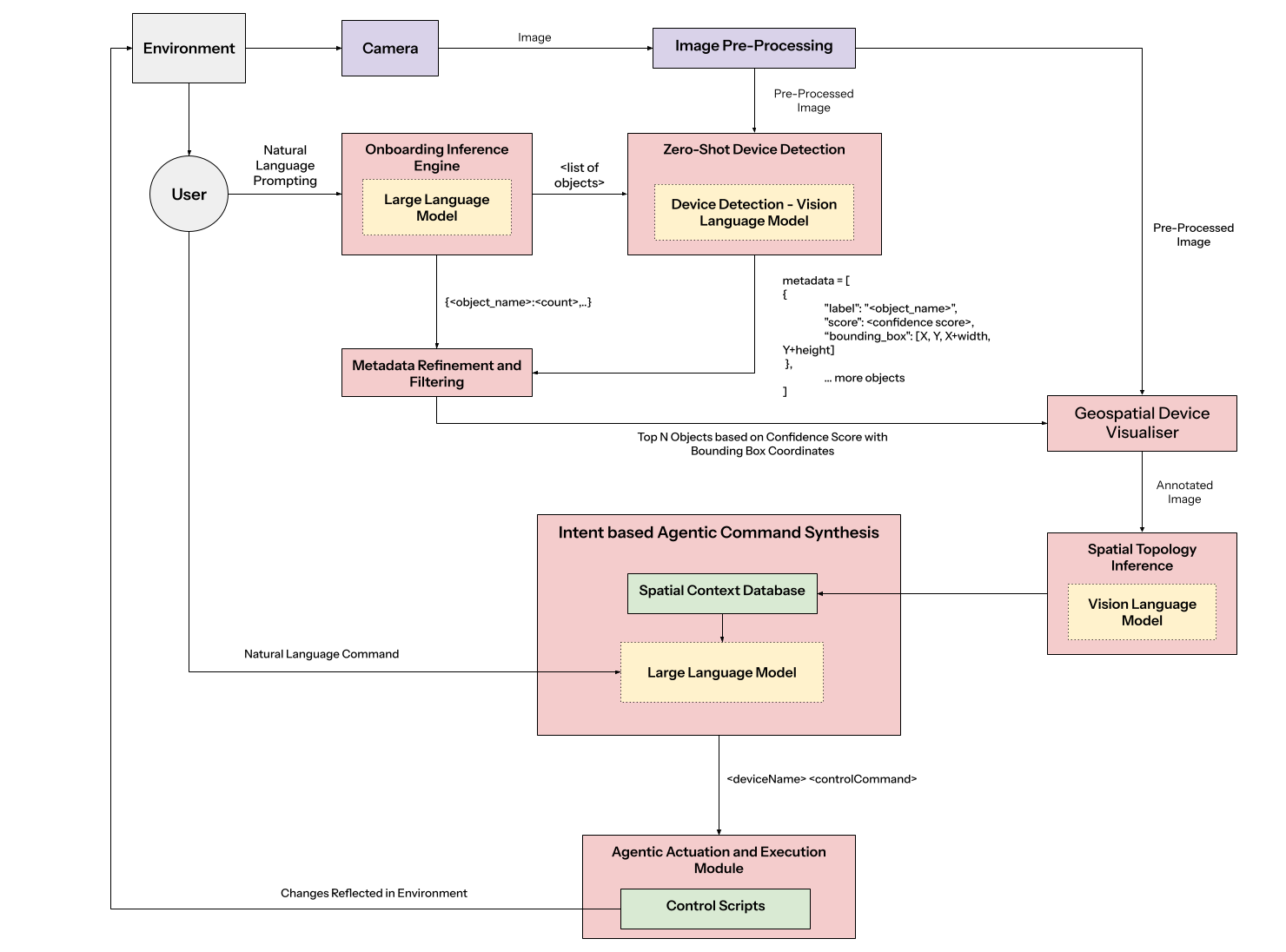}
    \caption{Overall Pipeline of the Proposed System}
    \label{fig:overall-pipeline}
\end{figure*}

The primary goal of the proposed system as shown in Figure \ref{fig:overall-pipeline} is to develop an intelligent, context-aware automation framework for smart home devices. By leveraging Vision Language Models (VLMs) and a modular architecture, the system ensures seamless interaction, precise control, and adaptive automation with minimal human intervention.

\textbf{Onboarding Inference Engine:} This module serves as the initial point of user interaction, collecting information about IoT devices present in the environment. It processes natural language inputs, enabling users to provide device details effortlessly. The extracted information is converted into a structured device inventory, which forms the basis for all subsequent modules.

\textbf{Zero-Shot Device Detection:} This module identifies and localizes IoT devices in a given image. Using OWL-ViT (Object-Aware Learning Vision Transformer), it performs zero-shot object detection, enabling the system to recognize previously unseen device types. The generated metadata provides vital attributes for each detected device, essential for precise control and automation.

\textbf{Metadata Refinement and Filtering:} To improve the accuracy of the system, this module processes the raw metadata generated by the detection module. It assigns unique identifiers and filters the data based on user inputs and model confidence scores. This ensures that only relevant and high-confidence detections are retained for further use.

\textbf{Geospatial Device Visualizer:} This component overlays bounding boxes and labels onto the input image based on the refined metadata. It provides users with an intuitive understanding of the device layout, supporting more effective automation decisions.

\textbf{Spatial Topology Inference:} This module analyzes the spatial configuration of devices by inferring their positions relative to room features and other IoT devices. Contextual spatial relationships are extracted to support intelligent automation strategies, ensuring optimal device coordination within the environment.

\textbf{Intent-Based Agentic Command Synthesis:} By combining spatial metadata with user intent, this module synthesizes precise control commands. It interprets real-time user instructions and environmental cues from the Spatial Topology Inference module to generate adaptive automation commands for responsive smart home interaction.

\textbf{Agentic Actuation \& Execution Module:} Serving as the final operational stage, this module interfaces with Tuya Smart Device API a smart home management platform. It executes the control commands generated by the system while handling validation and potential errors, ensuring smooth integration within the IoT ecosystem.

\vspace{1em}
\noindent\textbf{AI Models Utilized:}
\begin{itemize}
    \item \textbf{Qwen-2.5-32B:} Used for onboarding and interpreting natural language descriptions of IoT devices.
    \item \textbf{OWL-ViT (OWL2):} Responsible for automatic image-based annotation through zero-shot object detection.
    \item \textbf{GPT-4o:} Extracts spatial relationships and topology from annotated device data.
    \item \textbf{Gemini 2.0 Flash:} Processes user commands.
\end{itemize}

\noindent\textbf{Integration Platform:} \textit{Tuya Smart Device API} – used for executing control commands on real-world IoT devices.

\vspace{1em}

\subsection{Onboarding Inference Engine}
The Onboarding Inference Engine serves as the initial module in the Intelligence of Things control system. Its primary function is to facilitate user onboarding by collecting information regarding the number and types of IoT devices present in a given environment from the user. This ensures that the system is aware of the available devices before proceeding with subsequent detection and control processes.
\subsubsection{\textbf{Operational Mechanism}}
The operational mechanism of the Onboarding Inference Engine begins with user input collection, where the system prompts the user to provide details about the IoT devices present in their environment. This input can be provided in natural language and supports both text and voice modalities. For instance, users can state:
\texttt{
There are two fans and one light in this room.\\
One fan, three lights, and one air conditioner are present.\\
}

The system is designed to recognize various regional languages, including English, Tamil, Hindi, Telugu, and foreign languages as well. The input serves to specify the type and quantity of devices in the room, ensuring that the system can adapt to different user preferences and linguistic contexts.

Once the information is collected, the system processes the input to extract relevant device types (e.g., "fan", "light", "AC") and their corresponding quantities. This data is then structured into a standardized format to facilitate downstream processing. The structured data is passed to the next module (Zero-Shot Device Detection), ensuring that only relevant objects are considered for detection.

\subsubsection{\textbf{Prompt Used for Device Extraction}}
To effectively extract the number and type of IoT devices from user input, a predefined prompt is employed by the Onboarding Inference Engine. The prompt ensures that the system can consistently identify and quantify devices, regardless of the input format. The prompt is as follows:

\textbf{{Prompt:}
}\textit{You are an AI assistant responsible for onboarding users into a smart IoT control system. Your task is to extract the number and type of IoT devices mentioned by the user in natural language input.  
}

\textbf{{Rules:}  
}
\textit{\begin{enumerate}
  \item Identify the device type (e.g., \texttt{"fan"}, \texttt{"light"}, \texttt{"AC"}).
  \item Extract the quantity of each device.
  \item Ignore unrelated information and return only the structured device data.
  \item Store the output as a JSON dictionary with device types as keys and their counts as values.
\end{enumerate}
}

Example Input:
\textit{“There are 2 fans and 1 light in this room.” }\\

Expected Output:
\textit{ ["fan": 2,  "light": 1]}\\

This prompt is designed to handle a variety of user inputs and ensure that the extracted data are structured consistently.

\subsection{Zero-Shot Detection}
The Zero-Shot Device Detection Module is the foundational vision-based component responsible for identifying Internet of Things (IoT) devices from raw environmental imagery without prior task-specific training. This module uses OWL-ViT (OWL2)\cite{minderer2022simple}, an advanced zero-shot object detection model developed by Google, which enables detection of previously unseen object classes based on natural language prompts. By removing the need for retraining, OWL2 empowers the system to recognize a broad spectrum of smart devices in real-world environments.

\subsubsection{\textbf{Purpose and Role in the Pipeline}}
This module serves as the first vision-based inference step in the pipeline following the Onboarding Inference Engine (Section 4A), which provides a structured list of smart devices (e.g., fan, light, refrigerator, etc.) based on user input. The Zero-Shot Detection Module transforms these textual device types into object detection prompts and applies them directly on the input scene to locate the corresponding devices visually. The output from this module forms the raw annotated foundation for subsequent spatial reasoning and command generation.

\subsubsection{\textbf{Architecture and Functional Flow}}
The Zero-Shot Detection Module follows a deterministic multi-stage pipeline, as detailed below:

\paragraph{Device List Ingestion}
The system receives a predefined list of smart device classes, which are extracted based on the user’s initial instruction or onboarding scenario. This list consists of device types such as "fan," "light," "refrigerator," and others. Each device class is then converted into a natural language prompt for OWL2 inference. For example:
\textit{\begin{quote}
    \begin{itemize}
    \item “a ceiling fan”
    \item “a kitchen light”
    \item “a smart refrigerator”
\end{itemize}
\end{quote}}

\paragraph{Inference with OWL2}
OWL2 performs zero-shot object detection by embedding the visual features of the input image and comparing them against text-embedded representations of each device class prompt. The process is as follows:
\begin{itemize}
    \item For each device class prompt, OWL2 attempts to match the visual features of objects in the input scene with textual descriptions.
    \item For every successful match, the model produces:
        \begin{itemize}
            \item Bounding Box Coordinates: Defines spatial extent as   (x\textsubscript{1}, y\textsubscript{1}, x\textsubscript{2}, y\textsubscript{2}).
            \item Class Label: Device type (e.g., "fan," "light").
            \item Confidence Score: A value between 0 and 1.
        \end{itemize}
\end{itemize}

\paragraph{Metadata Structuring}
The raw detections are processed into a metadata format compatible with downstream modules. Each entry includes:
\begin{itemize}
    \item Device Type: Class label (e.g., "fan").
    \item Bounding Box: Coordinates (x\textsubscript{1}, y\textsubscript{1}, x\textsubscript{2}, y\textsubscript{2}).
    \item Confidence Score: Model confidence score.
\end{itemize}

All metadata entries are stored temporarily in a centralized file (`metadata.py`) for subsequent modules.

\subsection{Metadata Refinement and Filtering}
The Metadata Refinement and Filtering module is responsible for enhancing and structuring the raw detection metadata generated by the Zero-Shot Device Detection Module (Section 4B). This step ensures that only the most relevant and accurate device data is passed to the Geospatial Device Visualizer (Section 4D) and Spatial Topology Inference (Section 4E) modules. It processes the raw data by filtering out irrelevant detections, applying user-specific criteria, and generating clean metadata with consistent naming conventions and unique identifiers.

\subsubsection{\textbf{Purpose and Role in the Pipeline}}
This module acts as a quality control mechanism that refines the detection results from the OWL2 model. It ensures that the output passed to subsequent modules for visualization and spatial analysis is precise, relevant, and correctly labeled. The refinement process involves the assignment of unique identifiers \textbf{(UUIDs)} to each detected device, filtering out unwanted objects, and ranking detections by confidence scores.
The refined metadata provides structured, actionable data that enhances the system's spatial awareness and decision-making capabilities.

\subsubsection{\textbf{Key Functional Stages}}
The process of metadata refinement and filtering is carried out in the following stages:

\paragraph{Assigning Unique Identifiers (UUIDs)}
Each detected IoT device is assigned a Universally Unique Identifier (UUID) to ensure its consistency and trackability across different stages of the system. This unique ID links each device’s metadata to a specific instance, enabling precise tracking through the entire pipeline. UUIDs prevent confusion in environments with multiple devices of the same type (e.g., two lamps in the same room). This ensures that each device is treated as a distinct entity, even when identical devices are present in the same environment.

\begin{figure}[t]
    \centering
    \includegraphics[width=0.4\textwidth]{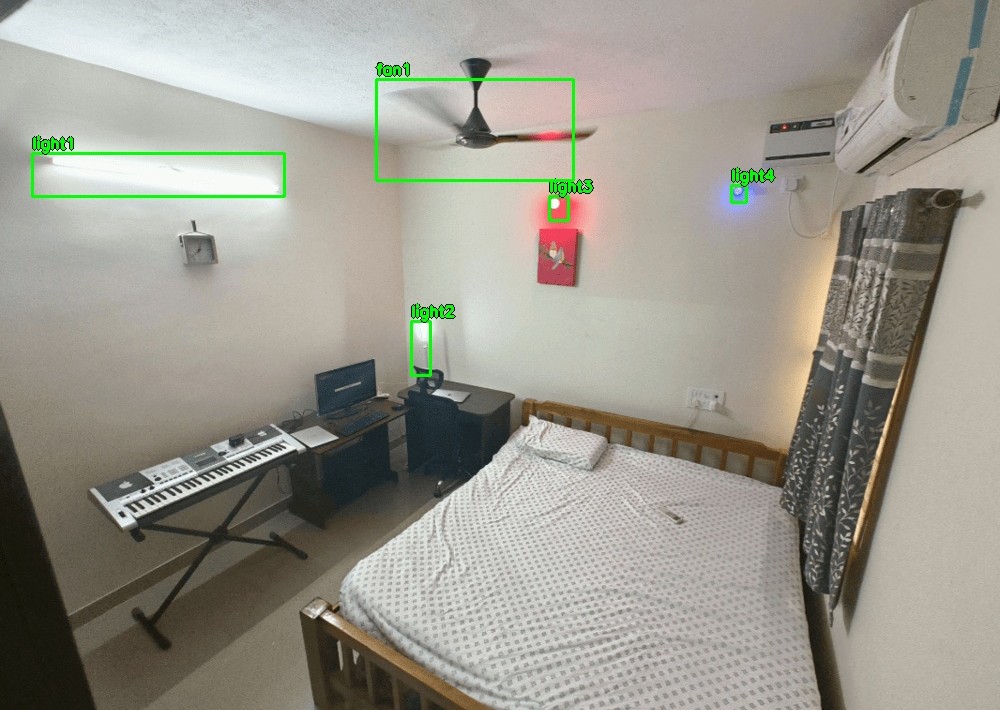}
    \caption{Annotations visualized by the Geospatial Device Visualizer module.}
    \label{fig:annotations}
\end{figure}

\begin{figure}[t]
    \centering
    \includegraphics[width=0.4\textwidth]{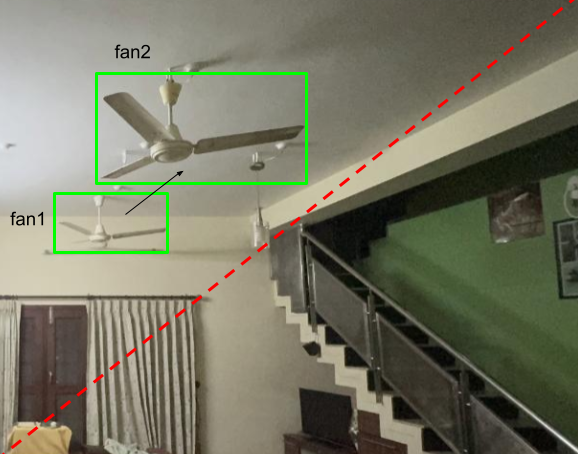}
    \caption{Labelling Nomenclature for Ambiguity Resolution}
    \label{fig:nomenclature}
\end{figure}

\paragraph{Naming Conventions for Detected Objects}
To maintain clarity and avoid ambiguity, the system adheres to a structured naming convention based on the spatial positioning of devices as shown in Figure \ref{fig:nomenclature} . The naming order is as follows:

\textbf{Horizontal Order:} Devices are named left to right based on their relative positions on the horizontal axis (e.g., from left to right across a room).
\textbf{Vertical Order:} When multiple devices of the same type are aligned on the same horizontal axis, they are named top to bottom, ensuring a consistent order.

This systematic approach helps ensure that all instances of the same device type (e.g., two lights) are uniquely identifiable based on their spatial location, minimizing the chance of confusion.

\paragraph{Filtering Devices Based on User Input}
The detected objects are compared against the user's initial command or Onboarding Inference Engine (Section 4A), which specifies the devices that the user wishes to control. The system filters out irrelevant devices that do not match the user’s request.

\paragraph{Ranking and Selecting Devices by Confidence Score}
The confidence score generated by OWL2 reflects the likelihood of an accurate detection. This module sorts the detected objects by their confidence score and selects the top-N most likely candidates for each device category.

\textbf{Example:} If the user requested two fans, the system selects the\textbf{ two fans with the highest confidence scores}, ensuring that only the most accurate detections are passed along.
\textbf{Thresholding:} A confidence threshold is applied to \textbf{discard detections with low confidence}, improving the reliability of the system’s outputs. For instance, detections with a confidence score below 0.5 may be excluded.

\paragraph{Outputting Refined Metadata}
After the filtering and ranking processes, the final metadata is structured in a clean JSON format containing relevant attributes:

 \textbf{Device Label:} e.g., fan, light, thermostat.
 \textbf{Bounding Box Coordinates:} e.g., (x\textsubscript{1}, y\textsubscript{1}, x\textsubscript{2}, y\textsubscript{2}), specifying the location of each object in the image.
 \textbf{Confidence Score:} e.g., 0.95, indicating the probability of a correct detection.
 \textbf{UUID}: Unique identifier for each device.

This refined metadata is then passed to the Geospatial Device Visualizer (Section 4D) and Spatial Topology Inference (Section 4E) for further processing.

\subsection{Geospatial Device Visualizer}
The Geospatial Device Visualizer module is responsible for translating refined metadata into a clear visual representation by drawing bounding boxes and labeling smart devices on the original image as shown in \ref{fig:annotations}. This step provides a human-interpretable snapshot of the detected environment and acts as an essential intermediary between raw detection and high-level spatial reasoning. By visualizing device locations spatially, the system not only confirms detection accuracy but also enables subsequent spatial relationship analysis in a meaningful and contextual manner.

\subsubsection{\textbf{Role in the Pipeline}}
This module acts as the visual bridge between object detection and spatial analysis. It consumes the filtered metadata from the Metadata Refinement and Filtering module (Section 4C) and overlays bounding boxes, labels, and identifiers directly on the input image. The resulting annotated image serves as the visual reference for the Spatial Topology Inference module (Section 4E), which analyzes the relative positions and spatial relationships between devices. In addition to enabling downstream AI processing, the visual output also serves as a validation and debugging tool for users and developers.

\subsubsection{\textbf{Functional Workflow}}
The module follows a structured process to convert refined metadata into a labeled, annotated image:

\paragraph{Loading the Image}
The input image, captured during the onboarding or command-triggered activation phase, is loaded using OpenCV. The image is then converted from BGR to RGB format to ensure accurate color representation and compatibility with image visualization tools.

\paragraph{Extracting Object Metadata}
The system retrieves metadata generated from the previous module, which includes the following attributes for each detected object:
\textbf{Device Type:} e.g., fan, light, TV.
\textbf{Bounding Box Coordinates:} (x1, y1, x2, y2).
\textbf{UUID:} A unique identifier for device tracking.
\textbf{Confidence Score:} Optional, used for annotation and filtering.
Objects are grouped by device type to aid in structured visualization and maintain semantic clarity.

\paragraph{Drawing Bounding Boxes and Labels}
For each device, a bounding box is drawn around its detected location using the provided coordinates. A text label is added to indicate the device type, and optionally, the UUID or confidence score. Distinct colors are dynamically assigned to each device type to visually distinguish them.

Label placement and bounding box styling are optimized for readability, non-overlap, and precise spatial representation.

\paragraph{Saving the Annotated Image}
The final annotated image is saved in a designated output directory in either PNG or JPG format. This image serves multiple purposes:
\begin{itemize}
    \item It is encoded in Base64 format and passed to the Spatial Topology Inference module (Section 4E) for AI-based spatial reasoning.
    \item It can be optionally displayed to the user or used by developers for debugging and validation of detection accuracy.
\end{itemize}

\paragraph{Error Tolerance and User Control}
To accommodate scenarios where the user is not satisfied with the automatic annotations, the system includes an error tolerance mechanism. Users have the ability to \textbf{refresh the automatic annotation process} triggering the automatic annotation pipeline to reprocess the image or manually \textbf{annotating the devices by drag-drop interface}. This allows for correction of inaccuracies, retrial of bounding box generation, and ultimately provides a more reliable visual outcome.

Through this visual annotation process, the system gains a spatially contextualized understanding of the environment, setting the stage for intelligent IoT control and interaction.

\begin{figure*}[t]
    \centering
    \includegraphics[width=\textwidth]{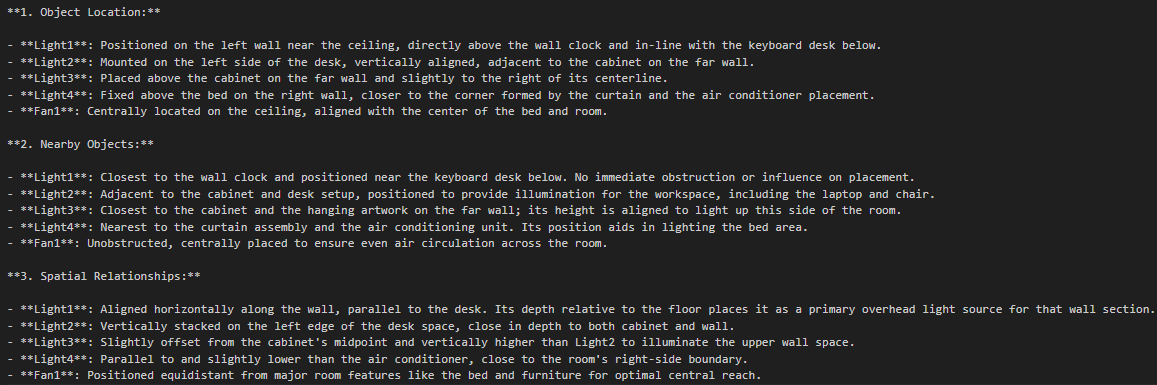}
    \caption{Output from the Spatial Topology Inference Engine}
    \label{fig:spatial-topology}
\end{figure*}

\subsection{Spatial Topology Inference Engine}
The Spatial Topology Inference module plays a pivotal role by analyzing the spatial configuration of detected IoT devices and their surrounding environment. Rather than treating smart devices as isolated entities, this module infers how devices relate to one another and to non-IoT environmental elements (e.g., furniture, walls, windows). These spatial insights allow the system to make intelligent decisions that consider both device placement and contextual constraints—paving the way for human-like spatial reasoning in smart environments.

\subsubsection{\textbf{Role in the Pipeline}}
This module acts as the critical bridge between visual perception and intelligent command generation. By leveraging GPT-4o\cite{hurst2024gpt} a vision language model, it transforms annotated images and structured metadata into meaningful spatial insights. It performs the following key roles:
\textbf{Interprets spatial layout:} Understands the arrangement and orientation of devices in the room.
\textbf{Enables context-aware commands:} Allows commands to factor in real-world constraints like furniture, visibility, and accessibility.
\textbf{Feeds decision-making engine:} Passes structured spatial descriptions to the Command Generation Module (Section 4F) to inform device targeting and reasoning.
This module immediately follows the Geospatial Device Visualizer (Section 4D) and serves as the final perceptual processing step before control logic is applied.
\subsubsection{\textbf{Functional Workflow}}
The Spatial Topology Inference module follows a multi-stage process:

\paragraph{\textbf{Input-} Annotated Image}
The input is the labeled image produced by the Geospatial Device Visualizer. All detected devices are enclosed in bounding boxes and annotated with class labels and UUIDs. This image is encoded in Base64 format for compatibility with the vision-language model.

\paragraph{\textbf{Input-} Object Metadata}
In parallel, the system provides GPT-4o with structured metadata, including:
\begin{itemize}
    \item \textbf{Device Types and UUIDs:} For device-level granularity.
    \item \textbf{Bounding Box Coordinates:} For relative positioning and size.
    \item \textbf{Confidence Scores:} For prioritizing more reliable detections.
\end{itemize}

\paragraph{Spatial Reasoning with GPT-4o}
The module dynamically generates a tailored prompt and sends it, along with the Base64-encoded image and metadata, to GPT-4o. The model analyzes the spatial layout and relationships using multimodal reasoning capabilities, producing a structured textual output.

\subsubsection{\textbf{Structured Prompt Design}}
To ensure focused and actionable results, the following structured prompt is used:

\begin{quote}
Analyze the provided image and the annotations \{object\_list\}(Module 1), ensuring the response does not exceed 1000 words.\\
\textbf{{Object Location:}} Briefly describe each device’s position relative to major room features (e.g., walls, windows, doors, furniture).\\
\textbf{{Nearby Objects:}} Identify the closest objects (IoT and non-IoT) and summarize their influence on placement.\\
\textbf{{Spatial Relationships:}
} Note relative depth, alignment, and positioning concisely, prioritizing only the most relevant details.
\end{quote}

This prompt structure ensures that GPT-4o:
\begin{itemize}
    \item Focuses on device layout and proximity to important environmental features.
    \item Avoids redundant or irrelevant commentary.
    \item Outputs structured, concise, and parsable text for downstream command generation.
\end{itemize}

By integrating advanced vision-language inference with spatial semantics as shown in Figure \ref{fig:spatial-topology}, this module equips the system with a high-level understanding of the environment—laying the groundwork for intelligent, context-sensitive control of smart devices.

\subsection{Intent Based Agentic Command Synthesis}
The Command Generation Module is the intelligence core of the IoT control pipeline that transforms spatially grounded device information and user intent into executable automation commands. This module fuses the natural language instructions provided by the user with the structured spatial metadata (from Section 4D), allowing the system to formulate contextually accurate, unambiguous, and actionable commands. It effectively bridges the gap between perception and control, enabling natural and intuitive smart-home interactions.

\subsubsection{\textbf{Role in the Pipeline}}
This module consumes outputs from:
\begin{itemize}
    \item \textbf{Onboarding Inference Engine} (Module 1): Provides user intent, utterance structure, and command semantics.
    \item \textbf{Spatial Topology Inference Engine} (Module 5): Supplies positional descriptions, contextual relationships, and disambiguation cues.
\end{itemize}

It produces executable and structured instructions for the \textbf{Execution Module}, enabling seamless device actuation.

\subsubsection{\textbf{Functional Pipeline}}
The Command Generation process follows a multi-stage approach:

\paragraph{Input Integration}
The system integrates multiple sources of contextual and perceptual information:
\begin{itemize}
    \item \textbf{User Command:} e.g., "Switch on the fan beside the window."
    \item \textbf{Device Metadata:} UUIDs and device types
    \item \textbf{Topological Descriptions:} e.g., Fan\_F2 is on the right wall, next to the window.
\end{itemize}

\begin{figure*}[t]
    \centering
    \includegraphics[width=\textwidth]{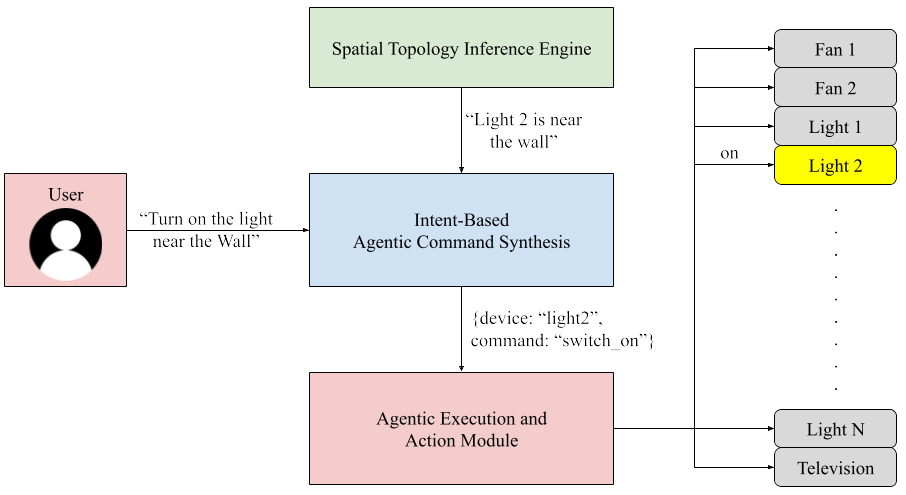}
    \caption{Command Flow for Execution}
    \label{fig:exec_comm_flow}
\end{figure*}

\paragraph{Prompt Construction for the Large Language Model (LLM)}
Gemini Flash \cite{team2024gemini} is chosen as the LLM for the module. To ensure high-quality command synthesis, this module constructs a structured prompt that integrates user intent and spatial metadata into a concise, instructive format suitable for Gemini. The design of this prompt plays a critical role in enabling the model to reason about spatial constraints and issue precise device commands.

The prompt construction process follows these key principles:

\begin{itemize}
    \item \textbf{Explicit Intent Embedding:} The user’s natural language command is clearly inserted at the beginning of the prompt to establish the task (e.g., “Turn on the fan near the window.”).
    
    \item \textbf{Device Enumeration with Spatial Context:} All detected devices are listed with:
    \begin{itemize}
        \item \texttt{UUIDs} for unambiguous referencing.
        \item Concise spatial descriptors derived from the Spatial Topology Inference Module (e.g., “left corner, near window”).
    \end{itemize}
    
    \item \textbf{Contextual Emphasis:} The prompt emphasizes critical spatial cues and relationships such as proximity, alignment, or room landmarks (e.g., “leftmost light,” “fan beside the door”).
    
    \item \textbf{Output Format Specification:} The prompt explicitly instructs Gemini to return output in a structured, parseable format using UUIDs and action keywords (e.g., \texttt{[UUID: <device\_uuid>, Action: <action>]}).
\end{itemize}

\textbf{Example Prompt to LLM:}
\begin{quote}
\textbf{{User Command:}} \textit{"Turn on the fan near the window."}

\textbf{{Device List:}}
UUID: fan-01, Position: left corner, near window.

UUID: fan-02, Position: center, far from window.
Based on the spatial information and user intent, identify the most appropriate device(s) and generate a control command in the following format:
[UUID: device-uuid, Action: action]

Ensure your response is concise and contextually accurate.
    
\end{quote}

This prompt formulation ensures that Gemini leverages both linguistic and spatial cues to resolve ambiguity and deliver accurate, targeted control instructions.

\paragraph{Command Structuring}
Gemini responds with a structured instruction in a machine-readable format such as JSON:
\begin{quote}
\begin{verbatim}
{
    "device": "light2",
    "command": "switch_on"
}
\end{verbatim}
    
\end{quote}

This standardized format ensures seamless parsing by the Execution Module for real-time actuation.

\paragraph{Handling Multiple Devices}
If the user command involves multiple devices (e.g., "Turn on the two lights on the left wall."), the module:
\begin{itemize}
    \item Filters candidate devices based on label, quantity, and spatial alignment.
    \item Selects the top-N matches by confidence score and relevance.
    \item Outputs a list of device-specific instructions.
\end{itemize}

By combining natural language understanding, spatial reasoning, and structured synthesis, this module enables a robust and intuitive interface for controlling smart environments in a human-like, context-sensitive manner.

\subsection{Agentic Execution and Action Module}
The Execution Module is the final component in the IoT control pipeline, responsible for taking the commands generated by the Command Generation Module and executing them on the appropriate smart devices. This module plays a crucial role in translating the synthesized commands into real-world actions, ensuring that the system’s automation functions seamlessly and effectively.

\subsubsection{\textbf{Input from the Command Generation Module}}
The Execution Module receives structured control commands from the Command Generation Module. These commands are typically in a JSON or dictionary format, specifying:
\begin{quote}
\textbf{UUID}: The unique identifier of the target IoT device.\\
\textbf{Action}: The specific action to be performed on the device (e.g., "turn-on", "adjust-brightness").
\end{quote}

\subsubsection{\textbf{Device Communication via TuyaAPI}}
Once the Execution Module receives the structured control commands, it communicates with the IoT devices via the TuyaAPI. TuyaAPI is a widely used platform that provides a standardized interface for controlling IoT devices over Wi-Fi. 
\paragraph{TuyaAPI Overview}
TuyaAPI allows for the seamless communication between the Execution Module and IoT devices. This API handles the underlying network communication, including connection establishment, message transmission, and response handling, thereby simplifying the process of interacting with devices over Wi-Fi.

\paragraph{Steps Involved in Communication}
\begin{enumerate}
    \item \textbf{API Authentication}: The first step involves authenticating the Execution Module with the TuyaAPI using an API key or other security tokens. This ensures that only authorized devices can communicate with the system.
    \item \textbf{Command Transmission}: After authentication, the Execution Module sends the command to the appropriate IoT device through the TuyaAPI. The API encodes the device's UUID and the corresponding action in the request.
    \item \textbf{Command Execution}: Upon receiving the command, the IoT device performs the requested action, such as turning on a fan or adjusting the brightness of a light. The device then sends a confirmation response back to the Execution Module.
    \item \textbf{Response Handling}: The Execution Module processes the response from the device to confirm the success or failure of the action. If successful, it logs the result and continues with the next action. If there is an error, the module can initiate error handling, including retries or user notifications.
\end{enumerate}

\paragraph{Example Communication Flow}
An example of a communication flow as shown in Figure \ref{fig:exec_comm_flow} between the Execution Module and a Tuya-enabled device is as follows:
\begin{itemize}
    \item The Execution Module generates a command to turn on the fan with UUID ``light\_02''.
    \item The command is transmitted to TuyaAPI, which forwards it to the smart light.
    \item The light receives the ``turn\_on'' command and powers on.
\end{itemize}

The Execution Module ensures that the IoT system operates as intended by translating commands into real-world actions. It achieves this by securely and reliably communicating with devices via the TuyaAPI, processing responses, and handling any potential errors to maintain consistent and efficient operation.

\section{User Study}
\label{sec:userstudy}
\subsection{Case Study Design}
Users are unaware of the smart device configurations and naming of devices in the environment to simulate a case where a user enters a new smart device home environment with no to minimal familiarity. A researcher demonstrates the usability of both Google Home and INOT assistants functionality to the users prior to the study to familiarise the technologies to users who have no experience of using smart home devices before.
For Google Home Assistant, the user decides on a task, then either refers to the map, asks the researcher for help, or recalls the unique ID label assigned to a specific device. They then issue a command to the Google Home Assistant AI using that device’s ID label, such as “Switch on light 4”.  For INOT assistant, the user decides on a task and directly issues a natural language command with spatial context, such as “Switch on the light that is near the AC,” without needing to refer to device IDs.
For consistency across trials, both systems were activated using designated hotkeys — the space bar for INOT and the microphone logo key for Google Home — to initiate voice interaction. Although voice-based activation using wake words was possible, it was deliberately avoided to reduce ambiguity and ensure uniformity across experimental conditions.
Reference tasks like the ones below were provided as examples, in case participants needed inspiration or guidance on the types of tasks that could be performed using both systems:
\textit{\begin{itemize}
    \item “You need to switch on the light near the AC.”
    \item “You need to switch on the light above the photo frame.”
    \item “You need to turn on the light on the desk.”
    \item “You need to switch on the leftmost light.”
    \item “You need to turn on the fan.”
    \item “You need to turn on the lighting for studying or working.”
\end{itemize}}

Participants were also encouraged to formulate their own open-ended commands based on their interpretation of the environment and the use cases they might want to explore.

\subsection{Participant Demographics}
Fifteen participants took part in the user study, with ages ranging from 18 to 80 years with a standard deviation (19.08) ensuring a diverse demographic range. The average age was 45.8 years, and the median age was 49. The gender distribution included 8 female participants (53.3\%) and 7 male participants (46.7\%). Educational backgrounds varied across the group: one participant had education below the 10th standard, another had completed senior secondary education (12th standard), and one participant held a doctoral degree. The remaining participants either held a bachelor's degree or were in the process of obtaining one.
Prior experience with smart home technologies was limited among the participants. Only two reported actively using smart devices, specifically Amazon Echo, in their homes. One participant mentioned possible past exposure, while the remaining participants had no prior experience with smart home systems.

Participants were briefed about the device ID labels in the environment only prior to the Google Home interaction session, to preserve the naturalism of the INOT experience. All required consents and ethical approvals were obtained before conducting the study.

\subsection{Material}
The experiments were conducted using a Google Home-compatible Android smartphone and a laptop running the INOT assistant with a working microphone. The laptop was equipped with an Intel® Core™ i7-10510U processor, 16 GB of RAM, and operated on the Windows 11 operating system.

\subsection{Results}
\subsubsection{\textbf{Experience and Usability}}
The ease of task completion was rated higher for INOT compared to Google Home Assistant. Google Home Assistant received a mean score of 3.8 and a median score of 4, while INTO Assistant scored a mean of 4.67 and a median of 4, on a scale where 1 indicated "very hard" and 5 indicated "very easy."
When asked about difficulties in expressing commands, most participants (6 out of 15) reported no issues. They indicated that, after becoming familiar with the system, they felt confident issuing commands without confusion. However, some participants experienced occasional challenges, particularly in the absence of INOT. These included difficulties in remembering device names when multiple devices were connected, confusion between "on" and "off" commands, and uncertainty regarding Google Home Assistant’s reliance on numerical identifiers, which sometimes made participants feel less confident during interactions.
Google Home Assistant was frequently noted as a source of confusion due to the use of numerical identifiers, which made users feel less certain. In contrast, INOT appeared to alleviate this confusion, allowing for more natural command expression.
Participants also reported occasional confusion with Google Home Assistant, especially in situations involving multiple devices or when distinguishing between similar-sounding commands. A few noted that the numerical system used by Google Home Assistant made them more self-conscious when issuing commands, which was not the case with INOT, where users were able to speak more freely and intuitively.

\subsubsection{\textbf{Emotional Reactions}}
Forty percent of users found Google Home Assistant easy and comfortable to use, but 53\% cited the need to remember device IDs as a major drawback. Additionally, three users mentioned that commands needed to be more specific. In total, 6 out of 15 users had a positive experience with Google Home Assistant, 4 had a neutral experience, and 5 had a negative experience.
In contrast, 73\% of users found INOT to be easy and intuitive, with no need to memorize device names being a major highlight. Users appreciated the spatial context-aware interaction and the use of regional languages, which made INOT more accessible. However, one user pointed out semantic ambiguity in object references, particularly with image-based identification (e.g., "Should I say photo, scenery, painting, or red board?"). Another user expressed dissatisfaction with the time constraints for issuing commands. Overall, 11 out of 15 users had a positive experience with INOT, 2 had a neutral experience, and 2 had a negative experience.
Eighty percent of users reported no confusion or frustration with either system. When confusion occurred, it was due to device ID dependency and multi-command processing in Google Home Assistant, and ambiguous references in INOT (e.g., object color/naming in images). Fourteen out of 15 users reported enjoying INOT, citing features such as automatic light mapping, easier complex command delivery, image-based spatial recognition, and robust natural language processing. Only one user preferred Google Home Assistant for its better mobile interface.

\subsubsection{\textbf{Preference and Future Use}}
14 out of 15 users preferred using INOT over Google Home Assistant, citing several key reasons. These included its ability to operate based on spatial context (e.g., “I would prefer INOT because it doesn’t control the light based on numbers”), ease of use, regional language support, and the lack of need for configuration. Users also appreciated INOT’s directional references, such as saying “turn on the lights to my left,” which worked seamlessly, unlike Google Home Assistant.
One user expressed privacy concerns about using INOT due to its processing of image data, although these concerns were noted as being addressed in the system. Another user preferred Google Home Assistant for its superior user interface.
Eighty percent (12/15) of users trusted INOT to control devices without needing to remember device names and IDs. Users also suggested improving INOT’s user interface and favored a more intuitive communication method with the system.

\subsubsection{\textbf{User Suggestions}}
Users provided additional suggestions not covered by prior questions. The most notable suggestions included implementing techniques to address privacy concerns, supporting all smart devices on the market, and ensuring that a phone can serve as one of the primary interfaces for INOT.

\begin{figure}[t]
    \centering
    \includegraphics[width=0.5\textwidth]{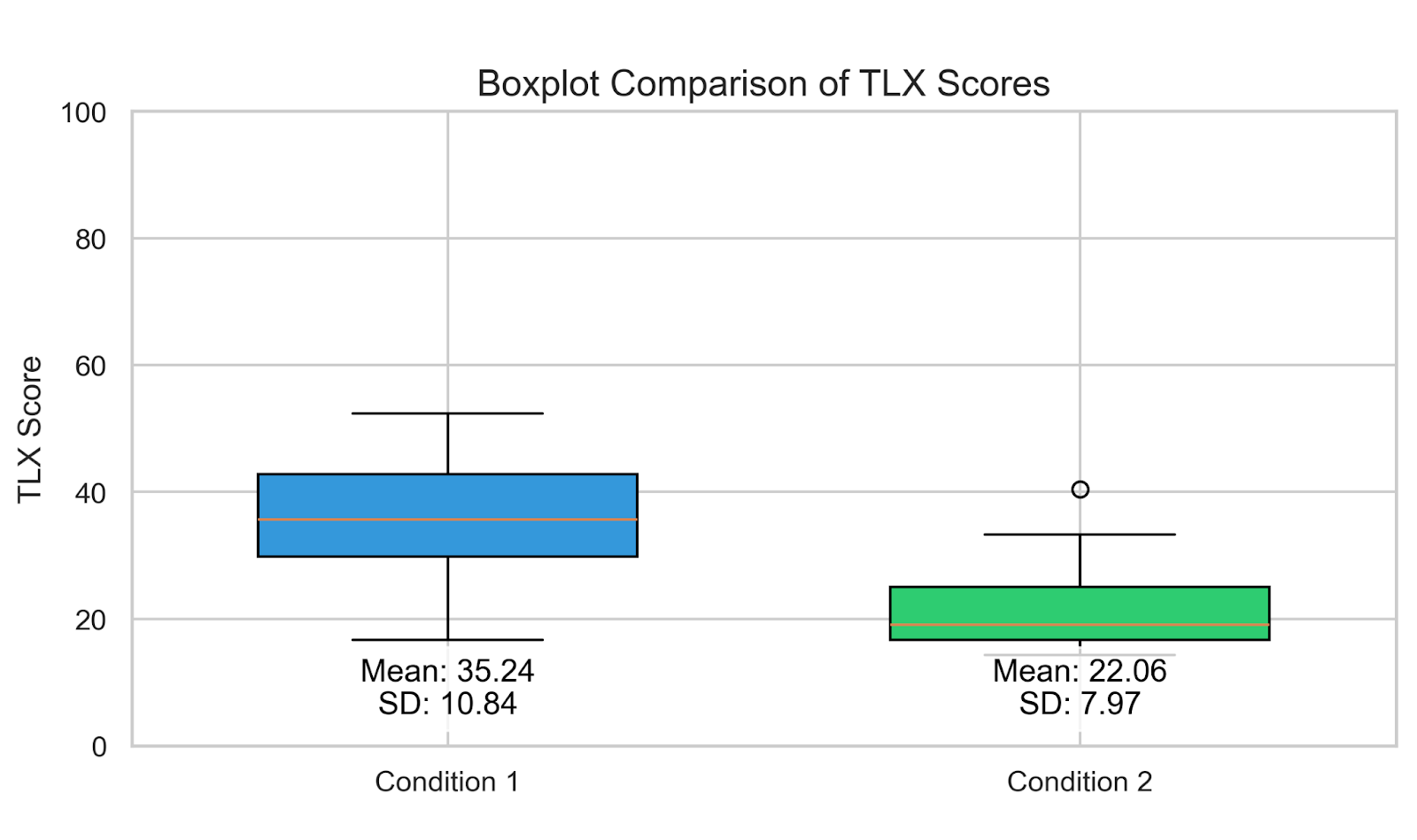}
    \caption{Condition 1 represents Google Home and Condition 2 represents INOT. The plot represents the generally lower TLX scores in Condition 2.}
    \label{fig:boxplot}
\end{figure}

\begin{figure}[t]
    \centering
    \includegraphics[width=0.5\textwidth]{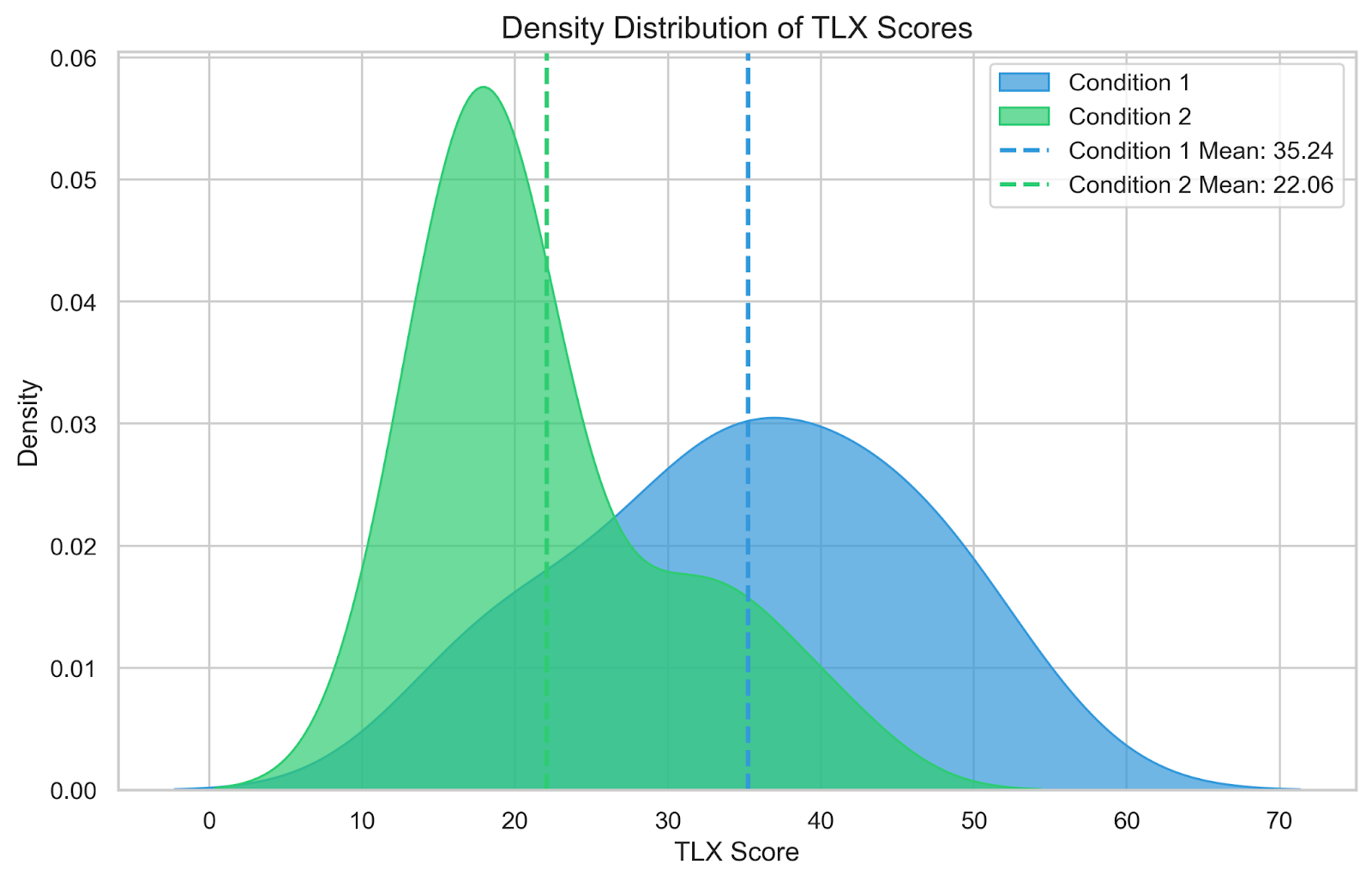}
    \caption{Density plot showing the distribution of scores for both conditions. \\Condition 1 (Google Home Assistant) is represented by Blue and Condition 2 (INOT Assistant) is represented by Green.}
    \label{fig:density_distribution}
\end{figure}

\subsubsection{\textbf{NASA Task Load Index (NASA-TLX) assessment}}
To evaluate the cognitive workload imposed by the baseline system - Google Home Assistant (Condition 1) and our proposed Intelligence of Things (INOT) system (Condition 2), we administered the NASA Task Load Index (NASA-TLX) to all participants after completing tasks in each condition.\\
\textbf{Descriptive statistics} indicate a substantial reduction in perceived workload when using the INOT system. The mean TLX score in Condition 1 was 35.24 (SD = 10.84), whereas the mean in Condition 2 dropped to 22.06 (SD = 7.97), representing an average decrease of 13.17 points. The median TLX also decreased from 35.71 to 19.05, and the interquartile range shifted lower across all percentiles, suggesting a broad and consistent improvement in user experience across participants.\\
\textbf{Inferential analysis} confirms that this reduction is statistically and practically significant. A paired-sample t-test revealed a significant difference between conditions (t(14) = 4.02, p = 0.0013). The corresponding Cohen’s d was 1.0381, indicating a large effect size. The 95\% confidence interval for the mean difference was [6.15, 20.20], confirming that the true effect is both statistically reliable and of meaningful magnitude. Normality of the differences was confirmed using the Shapiro-Wilk test (p = 0.4264), validating the use of parametric tests. Additionally, a Wilcoxon signed-rank test corroborated these results, revealing a significant difference (W = 10, p = 0.0045).\\
\textbf{Participant response patterns} further reinforce these findings. A total of 14 out of 15 participants (93.3\%) reported lower TLX scores in Condition 2, with no participants reporting identical scores between conditions. Only one participant (6.7\%) reported a higher TLX score in Condition 2, which may be attributed to individual variation or context-specific anomalies.\\
\textbf{Individual-level differences} ranged from modest reductions (e.g., 2.38 points) to substantial improvements exceeding 30 points, with the largest individual drop being 33.33 points. This strongly suggests that the INOT system consistently alleviates cognitive burden across diverse users and task scenarios.\\
In summary, these results provide compelling evidence that the INOT system significantly reduces perceived workload during smart-home interactions, outperforming the baseline condition with high consistency and large practical effect.

\section{Applications and Future Work}
\label{sec:applications}
\subsection{Smart Glasses Integration Using SLAM}
The integration of \textbf{smart glasses} with \textbf{SLAM (Simultaneous Localization and Mapping)} technology offers an advanced solution for \textit{hands-free interaction} with the environment. Smart glasses equipped with SLAM can map the environment in real time, allowing users to interact with the home automation system based on \textbf{spatial awareness}. For instance, users can issue commands such as, ``Turn on the light to my right,'' or ``Dim the lights in front of me,'' which dynamically change based on the user's orientation. This system enhances \textbf{user autonomy}, enabling more intuitive interactions without the need for \textit{touch-based devices} or static voice commands. Smart glasses, when coupled with \textit{spatial reasoning algorithms}, provide a versatile interface, particularly in contexts requiring continuous environmental awareness, such as for elderly individuals or those with disabilities. In addition, this integration facilitates \textit{hands-free operation}, improving ease of use in daily living environments.

\subsection{Installation in Smart Homes with Static Cameras}
The \textbf{INOT system} can be effectively deployed in residential smart home environments equipped with \textbf{static cameras}. By leveraging existing camera infrastructure—typically installed in common areas such as living rooms or kitchens—the system enables \textit{spatially-aware, natural language interaction} without requiring users to configure or memorize device identifiers. A single static camera per room is sufficient for INOT to perform \textit{real-time spatial reasoning}, \textit{object detection}, and \textit{natural language command interpretation}.

\subsection{Installation in Assisted Living Facilities}
The proposed system is highly suitable for \textbf{assisted living facilities}, where residents often require varying degrees of support for everyday tasks. The \textbf{natural language interface} and \textbf{spatially grounded commands} make the system especially beneficial in such contexts. Residents may face \textit{cognitive or physical challenges} that hinder traditional modes of interaction with smart devices. Unlike conventional smart assistants that depend on \textit{pre-configured labels} or \textit{app-based control}, INOT leverages \textbf{spatial reasoning} and \textbf{contextual awareness} to enable \textit{intuitive command formulation}—e.g., ``Turn on the fan next to the window'' or ``Switch on the light by the bed''—without requiring prior familiarity with the setup. This paradigm significantly reduces the \textit{cognitive and physical workload} involved in controlling home environments.

\subsubsection{Elderly Care and Independent Living}
The INOT system demonstrates significant potential for \textbf{elderly care} and \textbf{independent living} scenarios, where \textit{accessibility} and \textit{cognitive simplicity} are essential. Findings from our user study reinforce its utility in such contexts. Participants aged up to 80 years were included, with most having \textit{limited prior experience} with smart home technologies. Despite this, users consistently favored INOT over Google Home Assistant, citing its \textbf{intuitive, spatially-aware commands} and \textbf{regional language support} as major advantages. These features reduce reliance on memorizing device identifiers, a task particularly burdensome for \textit{older adults} or those experiencing \textit{cognitive decline}.

Furthermore, the significant reduction in \textbf{perceived cognitive workload}—evidenced by a mean \textbf{NASA-TLX score drop of 13.17 points} and a \textit{large effect size} (Cohen’s $d = 1.0381$)—indicates that INOT enables users to perform smart home tasks with \textit{less mental effort}. Participants also highlighted \textbf{emotional comfort} and \textbf{enjoyment} when using INOT, with \textbf{73\% reporting a positive experience}. The system's ability to interpret commands such as ``turn on the light near the photo frame'' or ``switch on the fan beside the bed'' without requiring \textit{exact terminology} enhances \textbf{accessibility} and supports \textbf{independent living}.

\subsubsection{Cognitive and Physical Impairment Support}
For users with \textbf{cognitive disorders} such as dementia, or those recovering from \textbf{neurological conditions} like strokes, remembering precise device names or operational sequences can be difficult. The INOT system addresses this by allowing users to refer to \textit{visual or spatial landmarks} rather than memorized terms. Similarly, for individuals with \textbf{motor impairments}—such as users in wheelchairs or those with \textit{Parkinson’s disease}—the system’s minimal reliance on \textbf{physical interaction} is particularly advantageous. Commands can be delivered \textit{verbally} from anywhere within the \textbf{field of view} of the static camera or smart glasses, further reducing the barrier to \textbf{accessibility}.

\section{Discussions and Future Work}
\label{sec:discussions}

\subsection{Developmental Insights}
The initial development of the \textbf{INOT system} began in a simulated environment using the \textit{Gazebo robotics simulator}. A virtual model of a domestic home was constructed to serve as the testing ground for early-stage prototypes. The first proof-of-concept employed a \textbf{tree-based spatial reasoning algorithm} that did not rely on deep learning techniques. In this version, spatial inference was handled through \textit{rule-based heuristics}, while a \textbf{large language model (LLM)} was responsible for generating commands for agentic execution.

To evaluate the practicality of this approach in more realistic scenarios, a \textbf{custom IoT simulator} was developed. This simulator modeled smart device actuation using \textit{signal indicators}, allowing experimentation with interaction paradigms without the need for physical hardware. However, the tree-based spatial reasoning algorithm demonstrated significant limitations in adaptability, \textit{depth analysis}, and \textit{generalization}—especially in complex, cluttered environments.

In response, the system was restructured to utilize \textbf{vision-language models (VLMs)} for spatial inference. VLM-based reasoning significantly improved performance, offering \textit{robust and context-sensitive} interpretations of spatial relationships from visual input.

Another critical development challenge involved the \textbf{reliable identification and referencing} of smart devices. Initial prototypes relied on \textit{manual annotation}, requiring users to label each device with a unique identifier—an approach that proved infeasible for large-scale deployment. The final solution integrated \textbf{computer vision algorithms} for automated device recognition, combined with a \textit{structured nomenclature system}, enabling scalable, context-aware interaction without manual intervention.

\subsection{Privacy Considerations}
While INOT enhances smart home usability through \textbf{continuous video-based spatial reasoning}, it inevitably raises \textit{privacy concerns}, particularly around visual monitoring of personal spaces. During the user study, one participant explicitly raised concerns regarding potential \textit{misuse or leakage of visual data}, especially in contexts involving persistent camera input.

Unlike systems that require training on user data, \textbf{INOT performs inference only} using \textit{pre-trained vision-language and language models}. It does not store or learn from user interactions, inherently reducing certain privacy risks. Nonetheless, the presence of always-on cameras presents valid concerns about surveillance, unintentional data capture, and sensitive information handling.

To address these, future work should explore \textbf{on-device processing}, ensuring that all inference tasks occur \textit{locally} without transmitting data to cloud servers. Additional techniques—such as \textbf{selective redaction} of sensitive visual regions, \textit{temporal deletion policies}, and \textbf{encrypted computation pipelines}—can further enhance user trust. Transparent communication about \textit{what data is processed, when, and how} will be essential to support ethical deployment. Privacy-preserving design must remain a \textbf{central focus}, especially in domains such as \textit{elder care and assisted living}.

\subsection{User-Centric Adaptation and Behavioral Modeling}
The core innovation of INOT lies in its \textbf{spatial context awareness}, but \textbf{personalization} is the next critical advancement. To further enhance the system, \textbf{user behavior modeling} should be prioritized. This includes recognizing individuals by their \textit{voice}, detecting \textit{emotional states} through vocal cues, and automatically adjusting device settings based on \textit{user preferences and moods}—all without requiring explicit commands.

Additionally, the integration of \textbf{rule-based automation}—such as \textit{"turn on the lights at 6 a.m."}—can provide intuitive, proactive assistance. To preserve user privacy while enabling learning, \textbf{federated learning} presents a promising direction, allowing models to improve \textit{locally} without centralized data collection.

As INOT scales across diverse households and behaviors, it must be capable of \textbf{adapting and evolving over time}. Focusing on \textit{model robustness and adaptability} in high-entropy domains will ensure that the system remains \textbf{responsive, personalized, and effective} as the user base grows and diversifies.

\section{Conclusion}
This research introduces the Intelligence of Things (INOT), a novel spatial context-aware control system for smart devices that fundamentally transforms how users interact with their IoT environments. By integrating advanced computer vision, natural language processing, and spatial reasoning capabilities, INOT addresses critical limitations in conventional IoT control systems that rely heavily on device-specific identifiers and pre-configured setups.
Our comprehensive user study demonstrated that INOT significantly outperforms traditional systems like Google Home Assistant across multiple metrics. The NASA-TLX assessment revealed a substantial reduction in cognitive workload with INOT (mean score of 22.06) compared to Google Home Assistant (mean score of 35.24), with 93.3\% of participants reporting lower cognitive burden. Qualitative feedback further reinforced these findings, with 87\% of participants preferring INOT due to its intuitive spatial context-aware commands, elimination of device ID memorization, and support for regional languages.
The modular architecture of INOT—comprising the Onboarding Inference Engine, Zero-Shot Device Detection, Metadata Refinement, Geospatial Device Visualizer, Spatial Topology Inference, and Intent-Based Command Synthesis—enables seamless adaptation to dynamic environments without requiring manual reconfiguration. This represents a significant advancement over existing solutions that typically require explicit device labeling and predefined automation rules.
While INOT demonstrates substantial potential, several challenges remain for future work. Enhancing privacy safeguards for image processing, expanding device compatibility across manufacturers, and optimizing the system for resource-constrained edge devices would further improve its applicability. Additionally, exploring lightweight LLM architectures specifically tailored to IoT environments could address computational limitations while maintaining real-time performance.
By bridging the gap between human spatial understanding and machine control, INOT paves the way for more intuitive, accessible, and context-aware smart environments. The system's ability to understand relative positions, interpret environmental contexts, and respond to natural language commands represents a significant step toward truly intelligent spaces that adapt to human needs rather than requiring humans to adapt to technological constraints.

\bibliography{mybib}

\end{document}